\documentclass[12pt,preprint]{aastex}

\newcommand\odt{$R_{23}$}
\newcommand{\tc}{$t^2$}

\newcommand{\ovsOH}{$R_{23}$ versus 12+log(O/H)}
\newcommand{\odtOH}{$R_{23}\,$--$\,$O/H diagram}
\newcommand{\Hiir}{\ion{H}{2} region}
\newcommand{\Hiirs}{\ion{H}{2} regions}

\newcommand{\HG}{HIIG}
\newcommand{\Hiigs}{\ion{H}{2} galaxies}

\newcommand{\told}{TOL $2146-391$}
\newcommand{\tolc}{TOL $0357-9315$}
\newcommand{\oiiim}{$T_e(4363/5007)$ method}
\newcommand{\oiiie}{4363/5007}

\shorttitle{\odt Recalibration}
\shortauthors{Pe\~na-Guerrero, Peimbert, \& Peimbert}

\received{}
\begin{document}

\title{RECALIBRATION OF PAGEL'S METHOD FOR \ion{H}{2} REGIONS CONSIDERING THE THERMAL STRUCTURE, THE IONIZATION STRUCTURE, AND THE DEPLETION OF O INTO DUST GRAINS}

\author{Mar\'ia A. Pe\~na-Guerrero\footnotemark[1,2]}
\email{guerrero@astro.unam.mx}

\author {Antonio Peimbert\footnotemark[1]}
\email{antonio@astro.unam.mx}

\and

\author{Manuel Peimbert\footnotemark[1]}
\email{peimbert@astro.unam.mx}

\footnotetext[1]{Instituto de Astronom\'ia, Universidad Nacional Aut\'onoma de M\'exico, Apdo. Postal 70-264, M\'exico 04510 D.F., Mexico}
\footnotetext[2]{Currently at Space Telescope Science Institute, 3700 San Martin Drive, Baltimore, MD 21218, USA; pena@stsci.edu}

\begin{abstract}
Using a sample of 28 \Hiirs\ from the literature with measured temperature inhomogeneity parameter, \tc, we present a statistical correction to the chemical abundances determined with the \oiiim. We used the \tc\ values to correct the oxygen gaseous abundances and consider the oxygen depletion into dust to calculate the total abundances for these objects. This correction is used to obtain a new calibration of Pagel's strong-line method, \odt, to determine oxygen abundances in \Hiirs. Our new calibration simultaneously considers the temperature structure, the ionization structure, and the fraction of oxygen depleted into dust grains. Previous calibrations in the literature have included one or two of these factors; this is the first time all three are taken into account. This recalibration conciliates the systematic differences among the temperatures found from different methods. We find that the total correction due to thermal inhomogeneities and dust depletion amounts to an increase in the O/H ratio of \Hiirs\ by factors of 1.7 to 2.2 (or 0.22 to 0.35 dex). This result has important implications in various areas of astrophysics such as the study of the higher end of the initial mass function, the star formation rate, and the mass-metallicity relation of galaxies, among others.
\end {abstract}


\section{Introduction}\label{intro}

The study of physical conditions of \Hiirs~is a fundamental tool to constrain models of nucleosynthesis of massive stars, galactic chemical evolution, and the chemical evolution of the Universe. In the same way, the study of the chemical composition of metal-poor \Hiirs\ is crucial for the understanding of the primordial chemical composition of the Universe. Most of these studies have been done using abundances obtained from the so called direct method, which assumes a homogeneous temperature throughout the whole volume of the object and then uses this temperature to determine abundances of all available ions; this temperature is commonly obtained from the [\ion{O}{3}] collisionally excited line ratio $\lambda \lambda\,$4363/(4959+5007); hereafter we will call this procedure the \oiiim.

Unfortunately, the simplification of the thermal structure of \Hiirs\ implied by the \oiiim, yields lower abundances than those obtained with other methods. For example: (i) the O/H abundances derived from recombination lines, (ii) the O/H abundances derived by comparing the nebular lines with photoionization models, and (iii) the comparison of the O/H nebular abundances (including gas and dust), with those derived from B stars and young F and G stars \citep[e.g.][]{ben06,kew08,prz08,pei11,sim11,nie12}. The abundances derived from these three different methods are consistent within the errors. The abundances obtained through collisionally excited lines corrected for the presence of thermal inhomogeneities are consistent with the abundances obtained through other methods. For reviews on such issue see \citet{pei11} and \citet{lop12}.

Abundances obtained with the \oiiim\ are underestimated because the temperature used in the determination process is assumed to be uniform. In the presence of either local or global thermal inhomogeneities, collisionally excited lines will be brighter in the hot regions while recombination lines will be brighter in the cool regions. This leads to choosing a temperature higher than the average and thus to assume that less oxygen (and more hydrogen), than the amount that is actually present, is needed in order to reproduce the observed line intensities. For this reason, these abundances should be corrected for the presence of thermal inhomogeneities before being used to calibrate any strong-line indicator. 

\citet{pag79} were the firsts to introduce the \ovsOH\ diagram (from here on, this diagram will be referred to as the \odtOH). They found that the sum of the intensities of nebular oxygen with respect to H$\beta$, {$I$([\ion{O}{2}]\,3727) + $I$([\ion{O}{3}]\,4959+5007)/$I$(H$\beta$)},
varied smoothly with the total oxygen abundance and proposed to use what is now known as $R_{23}$, $O_{23}$, or Pagel's method to determine the chemical abundances of objects where no auroral lines could be observed and, hence, no temperature could be determined. At present, more than a dozen strong-line indicators have been proposed; of them, the most studied is \odt\ \citep[e.g.][]{sta06,kew08,lop12}. These strong-line indicators can be calibrated using well known abundance determinations such as the ones based on the \oiiim\ or from fitting lines with numerical photoionization models. 

The quality of the abundances derived with the \oiiim, strongly depends on the validity of the assumption of constant temperature over the observed value. In the presence of large thermal inhomogeneities, the \oiiim\ will underestimate abundances since the temperature dependence of the intensity of collisionally excited lines (such as [\ion{O}{3}] $\lambda \lambda$ 4363, 4959, and 5007) is very different from that of the recombination line H$\beta$. The intensity of collisionally excited lines is proportional to $I\varpropto T_e^{-1/2} \times exp(-\Delta~E/$k$ T_e)$, while the intensity of H$\beta$ is proportional to $I\varpropto T_e^{-0.87}$, where $T_e$ is the local electron temperature, k is the Boltzmann constant, and $\Delta$~E is the collisional energy required to excite the line. As an alternative to avoid the problem of the large temperature dependence of collisionally excited lines, recombination lines can be used to determine abundances and calibrate the \odtOH\ because they have a temperature dependence given approximately by $I\varpropto T_e^{-1}$. For recombination line to H$\beta$ ratios, the temperature dependence almost cancels out, whereas for collisionally excited line to H$\beta$ ratios the temperature dependence does not cancel out. Some work has been done in the upper branch of the diagram using recombination lines, however, these faint lines can only be observed in metal-rich objects. 

The discrepancy between abundances determined using recombination lines and collisionally excited lines is called the abundance discrepancy factor, ADF, problem. Typical values for \Hiirs\ lie in the 1.5 to 3 range \citep[][and references therein]{pei93,pea05,gar07,pei07,est09}. There have been two major explanations for this discrepancy: (i) high metallicity inclusions that will create cool high density regions surrounded by hot low density regions \citep[e.g.][]{tsa05}, and (ii) thermal inhomogeneities in a chemically homogeneous medium that are caused by various physical processes such as shadowed regions, advancing ionization fronts, shock waves, magnetic reconnection, etc. \citep[e.g.][]{pei11}. Since the densities determined from oxygen recombination lines correlate very well with the [\ion{Cl}{3}] densities \citep{pea05a}, we prefer the second explanation.

In Section \ref{tempinhom} we discuss the corrections due to thermal inhomogeneities and due to the fraction of O depleted into dust grains. In Section \ref{CDM} we present a quantitative correction to the \oiiim, this correction is applied to the \odt\ method in Section \ref{recal}. In Section \ref{disc} we discuss the relation between the O$^{++}$ fraction and the intensity of the [\ion{O}{2}] and [\ion{O}{3}] nebular lines. Finally, in Section \ref{conc} we present the conclusions.

\section{Temperature Inhomogeneities and Dust}\label{tempinhom}

The formalism to study the chemical abundances in the presence of thermal inhomogeneities was introduced by \citet{pei67} and \citet{pei69}, where the parameter that characterizes such thermal inhomogeneities was defined as \tc. In these works it was recognized that thermal inhomogeneities would produce systematic differences in temperatures determined from different methods, and that abundances determined with each one of those temperatures would also present systematic differences. The abundances determined using the \oiiim\ were approximately a factor of 2.5 lower than those determined using the same collsionally excited line intensities including the effect of thermal inhomogeneities.

There are several ways to obtain a \tc\ value, six of them are discussed in \citet{pei11}. These are based on the comparison of: (i) the O$^{++}$ abundances obtained using [\ion{O}{3}] lines versus abundances obtained using \ion{O}{2} lines; (ii) the temperature derived from \ion{He}{1} lines versus the temperatures derived from [\ion{O}{2}] and [\ion{O}{3}] lines; (iii) the temperatures derived from the ratio of the Balmer and Paschen continua to the Balmer line intensities versus the temperatures derived from [\ion{O}{2}] and [\ion{O}{3}] lines; (iv) the C$^{++}$ abundances derived using [\ion{C}{3}] and \ion{C}{3}] lines versus abundances obtained from \ion{C}{2} lines; (v) 1.5$\times10^6$ columnar temperatures determined from a high spatial resolution map of the Orion nebula \citep{ode03}; and (vi) the calibrations of the \odt\ method using observations versus those using photoionization models. In all cases, the \tc~values are consistent with each other, which implies that those objects are chemically homogeneous. The first two methods are the most used to determine \tc.

Independently of the effects of the thermal structure, there is an additional correction due to dust depletion that should be taken into account to determine abundances. \citet{est98} found that the depletion of oxygen into dust grains in the Orion nebula amounted to a 0.08 dex correction in the O/H ratio; slightly higher estimates of this correction have been found in recent studies of the Orion nebula \citep{mes09b,sim11}. From the depletion of Mg, Si, and Fe in Galactic and extragalactic \Hiirs, \citet{pea10} estimated an O/H correction ranging from 0.09 to 0.11 dex that increases with increasing O/H ratios. We have used the correction recommended by Peimbert et al. for all the objects of our sample; this corrections amounts to 0.11 dex for those objects whose gaseous oxygen abundance is 12+log(O/H)$>8.3$, 0.10 dex for objects with $8.3>$12+log(O/H)$>7.8$, and 0.09 for those objects with $7.8>$12+log(O/H).

\section{Correction to the abundances derived with the \oiiim}\label{CDM}

Due to the inconsistencies presented when comparing abundances obtained with the \oiiim\ and those obtained with, recombination lines, photoionization models, and using abundances of recently formed stars; it is clear that abundances derived from the \oiiim\ are systematically underestimated. This needs to be corrected by considering the thermal structure of the object and the fraction of oxygen depleted into dust grains. 

Table \ref{trec}\ lists the sample we used to correct the \oiiim. The sample consists on six \Hiigs, eight Galactic \Hiirs, and 14 extragalactic \Hiirs. This total of 28 objects were gathered from the literature and, in each of them, at least one value of \tc\ was obtained. For the cases where there was more than one \tc\ value we took the one presented as average or the one from the brightest observed region. 

Based on detailed analysis presented in the literature it follows that the total oxygen abundance should be higher due to the effects of the thermal structure on the emitted spectra \citep[e.g.][]{pei67,pei69,pei00,pea03,gar04,est05,pei07,est09,pei11,pen12,pea12} and that the fraction of O depleted into dust grains should be taken into account \citep[][see also section \ref{tempinhom}]{pea10}. We expect the magnitude of these effects to depend on the oxygen abundance and, possibly, on the oxygen excitation degree. In order to take these factors into account, we considered that the abundances derived from the \oiiim\ should be corrected with a function $f$ dependent on the oxygen ionization degree and on the oxygen abundance determined from the \oiiim, (O/H)$_{\oiiie}$:
\begin{equation}
{\rm (O/H)_{CALM}} = {\rm (O/H)_{\oiiie}} + f( P, {\rm (O/H)_{\oiiie}}),
\label{corrtipo}
\end{equation}
where CALM stands for Corrected Auroral Line Method. Throughout this paper all oxygen abundances are given in units of 12+log(O/H).

 As a first approximation to the function $f$, we decided to use a linear correction of the form 
\begin{equation}
 f(P, {\rm (O/H)_{\oiiie}}) =  C_1+C_2 \times  {\rm (O/H)_{\oiiie}} +
C_3  \times P + C_4  \times P  \times {\rm (O/H)_{\oiiie}},
\label{corrf}
\end{equation}
where $C_1$, $C_2$, $C_3$, and $C_4$ are constants.

All abundances determined for our sample were corrected for (i) the presence of thermal inhomogeneities and (ii) the fraction of oxygen depleted into dust grains. We have found that the difference between these corrected abundances and those determined with the \oiiim\ lie in the 0.16 to 0.40 dex range. 

\begin{figure}
\begin{center}
\includegraphics[angle=0,scale=0.7]{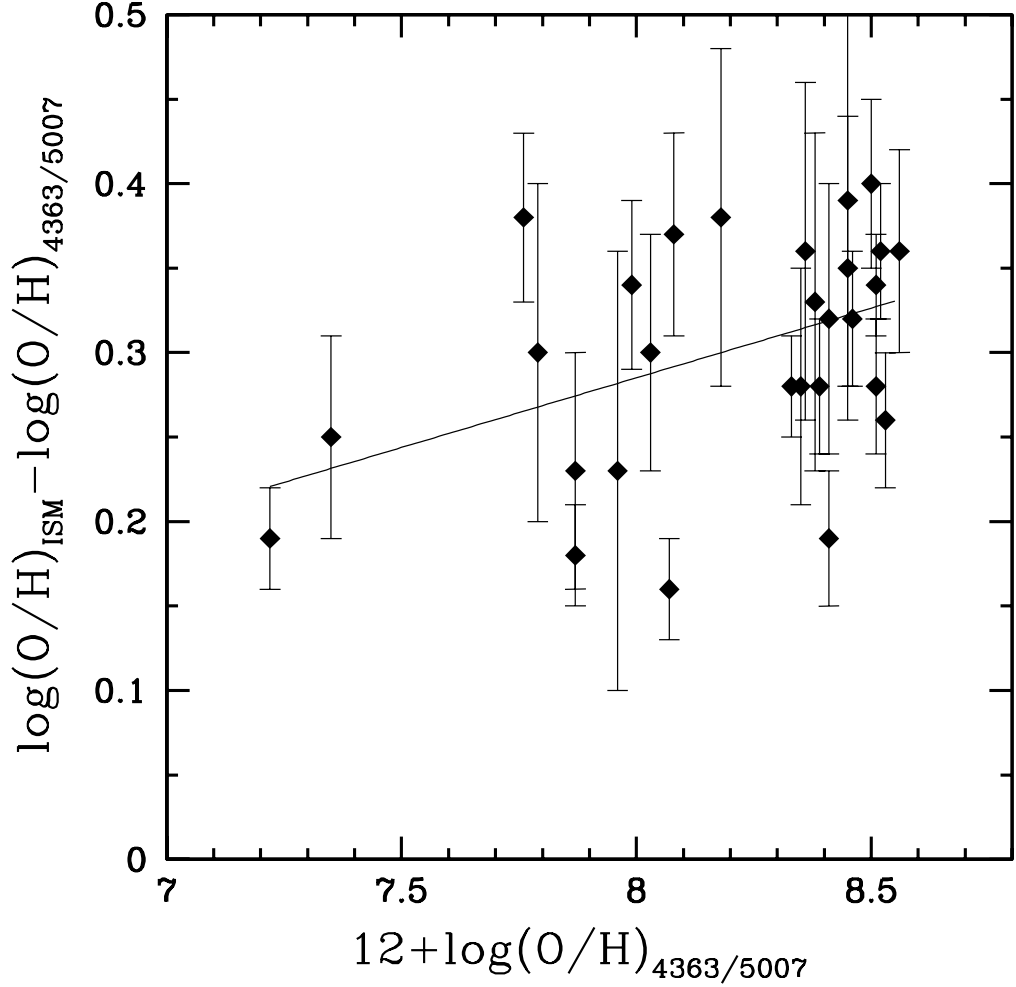}\\
\caption{Plot of the abundances obtained with the \oiiim, versus the difference in dex between the corrected abundances and those obtained with the \oiiim. The line is the best linear fit to the objects presented in Table \ref{trec}, log${\rm (O/H)_{ISM}}-$log(${\rm O/H_{\oiiie}}$)$=0.0825\times$[12+log(${\rm O/H_{\oiiie}}$)]$-$0.375.
\label{ecrecal}}
\end{center}
\end{figure}

We found very little dependence with oxygen degree of ionization: $C_3$ and $C_4$ are $0.010$ and $0.0002$, respectively. Since the dependence with ionization degree is small and consistent with zero we will ignore it. Figure \ref{ecrecal} presents the oxygen abundance correction dependence with oxygen abundance for the objects of our sample. The best fit is obtained with $C_1$ and $C_2$ equal to $-0.375$ and $0.0825$, respectively:
\begin{equation}
f(P, {\rm (O/H)_{\oiiie})}=f({\rm (O/H)_{\oiiie}})=0.0825 {\rm (O/H)_{\oiiie}}-0.375;
\label{corrlineal}
\end{equation}
this implies a corrected oxygen abundance:
\begin{equation}
{\rm (O/H)_{CALM}}=1.0825{\rm (O/H)_{\oiiie}}-0.375.
\label{DMC}
\end{equation}
The points in Figure \ref{ecrecal} show considerable dispersion; this is not surprising, because (i) the errors in the determination of each \tc\ value are large and (ii) the true value of \tc\ for each object depends on the details of the object \citep[see Figure 4 of][]{pea12}. To test for the statistical significance of this correlation we calculated a Spearman's correlation statistic of $\rho=0.3685$, this suggests that the correlation is real with 95\% confidence. When comparing the dispersion with the uncertainties on the oxygen abundance determinations for this sample we calculated $\chi^2=60.28$; showing that about half of this dispersion comes from the uncertainties of the determination of \tc; meaning that approximately half of the dispersion comes from the unique physical conditions of each object. 

\section{Recalibration of the \odt\ Abundance Determination Method}\label{recal}

\odt\ is now being widely used to determine abundances in objects with low intrinsic brightness or objects with low to intermediate redshift ($z$ up to 1).
The \odt\ method remains one of the most popular strong line indicators because (i) it is based directly on oxygen lines, (ii) it includes lines from the 2 relevant ionization stages, (iii) the lines it uses are the most intense for most \Hiirs, and (iv) these lines lie in the blue part of the spectrum and can be observed from the ground for higher redshifts than other methods.

\begin{figure}
\begin{center}
\includegraphics[angle=0,scale=0.7]{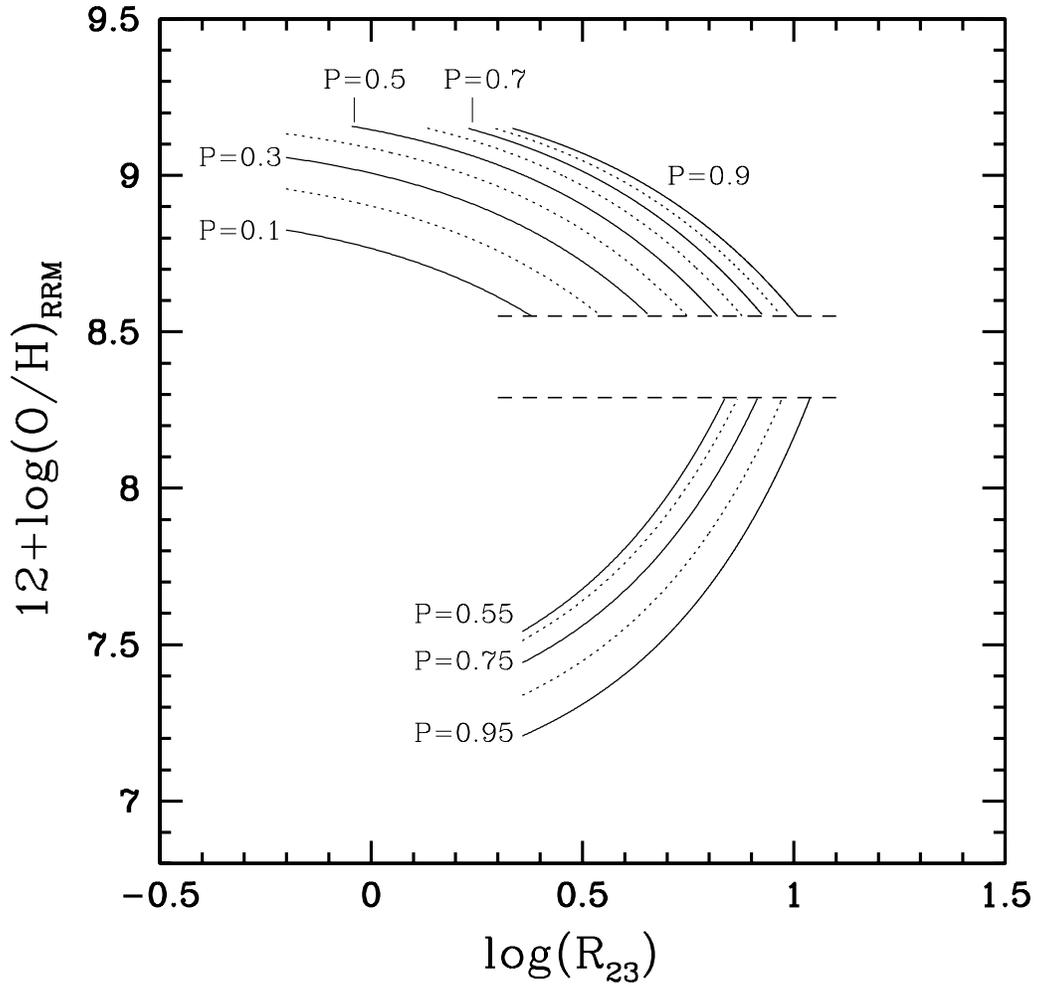}\\
\caption{New calibration of the \odt\ Method from equations \ref{recA} and \ref{recB}. For clarity, the solid $P$-valued curves have been labeled. The dashed horizontal lines indicate the transition zone between the upper and lower branches.
\label{figrecal}}
\end{center}
\end{figure}

Unfortunately, the curve in the \odtOH\ is double valued (see Figure \ref{figrecal}); this occurs  because: at low metallicities, the main cooler in \Hiirs\ is the hydrogen atom; with increasing element abundances, the cooling due to oxygen nebular lines increases, hence, the value of \odt\ increases reaching a maximum value of \odt$\,\approx10$ when 12+log(O/H)$_{\rm ISM}\approx8.4$; then, at higher metallicities, the IR [\ion{O}{3}], IR [\ion{Ne}{2}], and optical [\ion{N}{2}] lines begin to dominate the cooling so that the total value of \odt\ decreases as the metallicity increases \citep{van98}. 

Also, there is a noticeable dependence on the oxygen ionization degree \citep[e.g.][]{mcg91,kew02}. In a series of articles \citet{pil00,pil01a} and \citet{pil05}, presented an analytic family of curves for each branch of the \odtOH. This family of curves depends on the oxygen excitation degree across the object; defining the $P$ parameter as the observable oxygen excitation ratio, $P$=([\ion{O}{3}] 4959+5007)/([\ion{O}{2}] 3727 + [\ion{O}{3}] 4959+5007), and presenting two equations to determine the upper and lower branch abundances as a function of (O/H)$_{\rm P}$=$f$($P$,\odt). This calibration is based on abundances determined using the \oiiim. 

Unfortunately, depending on the calibration, resultant O/H values can vary by as much as 0.6 dex \citep{pei07,kew08}. This large dispersion in the calibrations mainly comes from what was chosen to be adjusted: (i) the theoretical intensities given by photoionization models, (ii) abundances derived from the \oiiim, or (iii) abundances derived from observations of oxygen recombination lines. The lowest determinations for the gaseous oxygen abundances are those that come from using the \oiiim, therefore they require the correction due to thermal inhomogeneities; additionally, to obtain the total oxygen abundances, a correction of $\sim0.10$ dex ---due O depletion into dust grains--- is required.

We find that, among all the calibrations of the \odtOH\ available in the literature, the one presented by \citet{pil05} is the most adequate for this work since it explicitly includes the oxygen ionization degree through the $P$ parameter.

The calibration of \citet{pil05}, (O/H)$_{\rm P}$, used abundances that were determined with the \oiiim; hence, abundances are underestimated but this allows us to use equation \ref{DMC} to correct them since (O/H)$_{\rm P}$ has abundances determined with the \oiiim\ at its core. We can determine oxygen abundances from a recalibration of the \odt\ method, (O/H)$_{\rm RRM}$, by substituting (O/H)$_{\rm RRM}$ and (O/H)$_{\rm P}$ in place of (O/H)$_{\rm CALM}$ and (O/H)$_{\oiiie}$ into equation (\ref{DMC}). We take (O/H)$_{\rm P}$ from equation  [22] and [24] of Pilyugin and Thuan for the upper and lower branches, respectively, to obtain two recalibrated families of curves that now simultaneously consider the thermal structure, the ionization structure, and the depleted fraction of O into dust grains. The oxygen abundance for the upper branch (which shifted from 12+log(O/H)$_{\rm P}\geq$8.25 up to 12+log(O/H)$_{\rm RRM}\geq$8.55), can be estimated from:\\
\begin{eqnarray}
{\rm (O/H)_{RRM}}&=&\left( \frac{R_{23}+726.1+8.42P+327.5P^2}
{85.96+82.76 P+43.98P^2+1.793R_{23}}\right) 1.0825-0.375 \nonumber \\
&=&\frac{R_{23}+1837+2146P+850P^2}
{209.5+201.7 P+107.2P^2+4.37 R_{23}},
\label{recA}
\end{eqnarray}
while the oxygen abundance for the lower branch (which shifted from 12+log(O/H)$_{\rm P}\leq$8.00 up to 12+log(O/H)$_{\rm RRM}\leq$8.29), can be estimated from:\\
\begin{eqnarray}
{\rm (O/H)_{RRM}}&=&\left( \frac{R_{23}+106.4+106.8P-3.40P^2}
{17.72+6.60P+6.95P^2-0.302 R_{23}} \right) 1.0825-0.375 \nonumber \\
&=& \frac{R_{23}+90.73+94.58 P-5.26P^2}
{14.81+5.52 P+5.81 P^2-0.252 R_{23}}.
\label{recB}
\end{eqnarray} 

\begin{figure}
\begin{center}
\includegraphics[angle=0,scale=0.7]{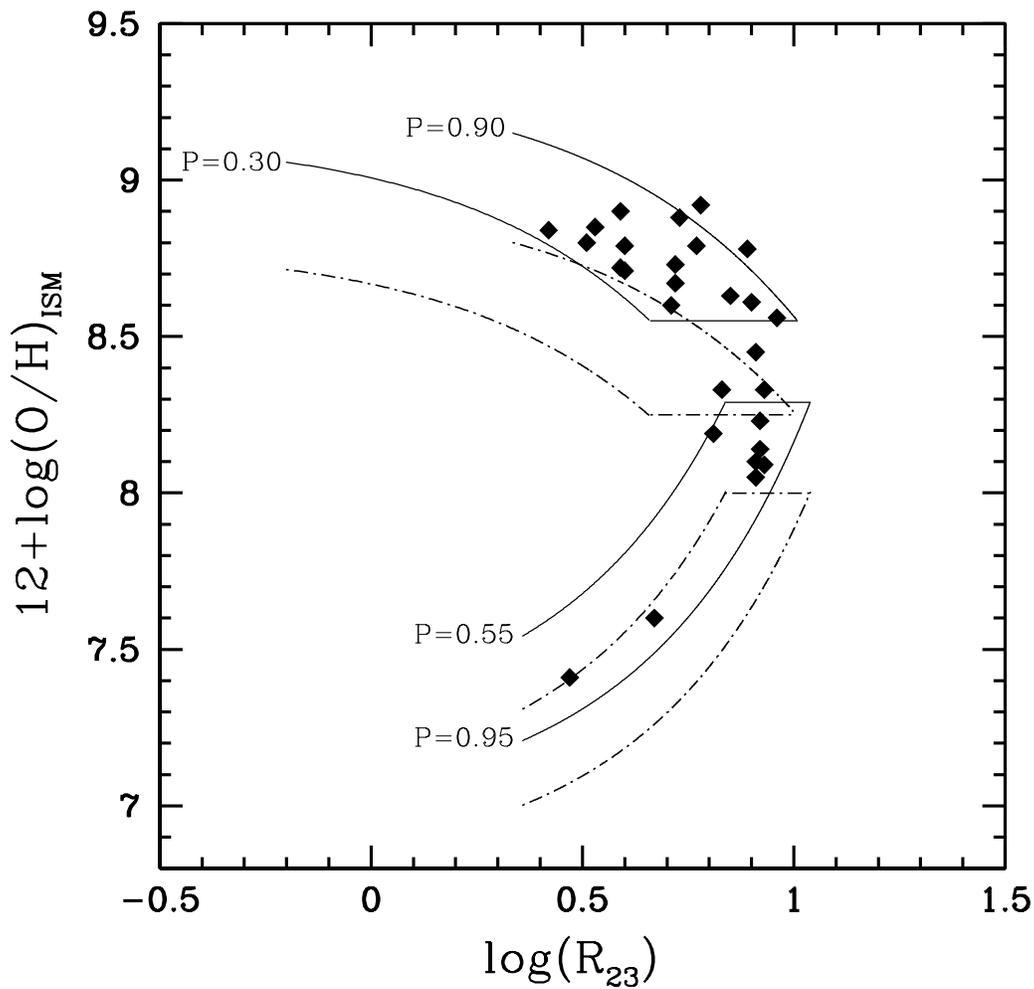}\\
\caption{ Plot of the oxygen abundance vs. \odt\ for the 28 objects from Table \ref{trec}, where abundances were determined considering the presence of thermal inhomogeneities as well as the fraction of O depleted into dust grains. For comparison we present representative bands of the upper and lower branches for both nebular line calibrations, (O/H)$_{\rm RRM}$ (solid black curves) and (O/H)$_{\rm P}$ (point-dashed curves). 
\label{frecalpt}}
\end{center}
\end{figure}

Figure \ref{figrecal} shows these recalibrations of the upper and lower branches for different oxygen ionization degrees, while Figure \ref{frecalpt} compares the calibration presented in this work and that of \citet{pil05} to the data from Table \ref{trec}.

\section{The O$^{++}$ fraction}\label{disc}

Something missing when using the $P$ parameter is a quantitative connection with the O$^{++}$ fraction in the nebulae. We will define the O$^{++}$ ionization degree as:
\begin{equation}
{\rm OID}=\frac{\int n_e n( {\rm O^{++}}) dr}{\int n_e  n({\rm O^{+}+O^{++}}) dr}.
\label{eqoid}
\end{equation}

\begin{figure}
\begin{center}
\includegraphics[angle=0,scale=0.7]{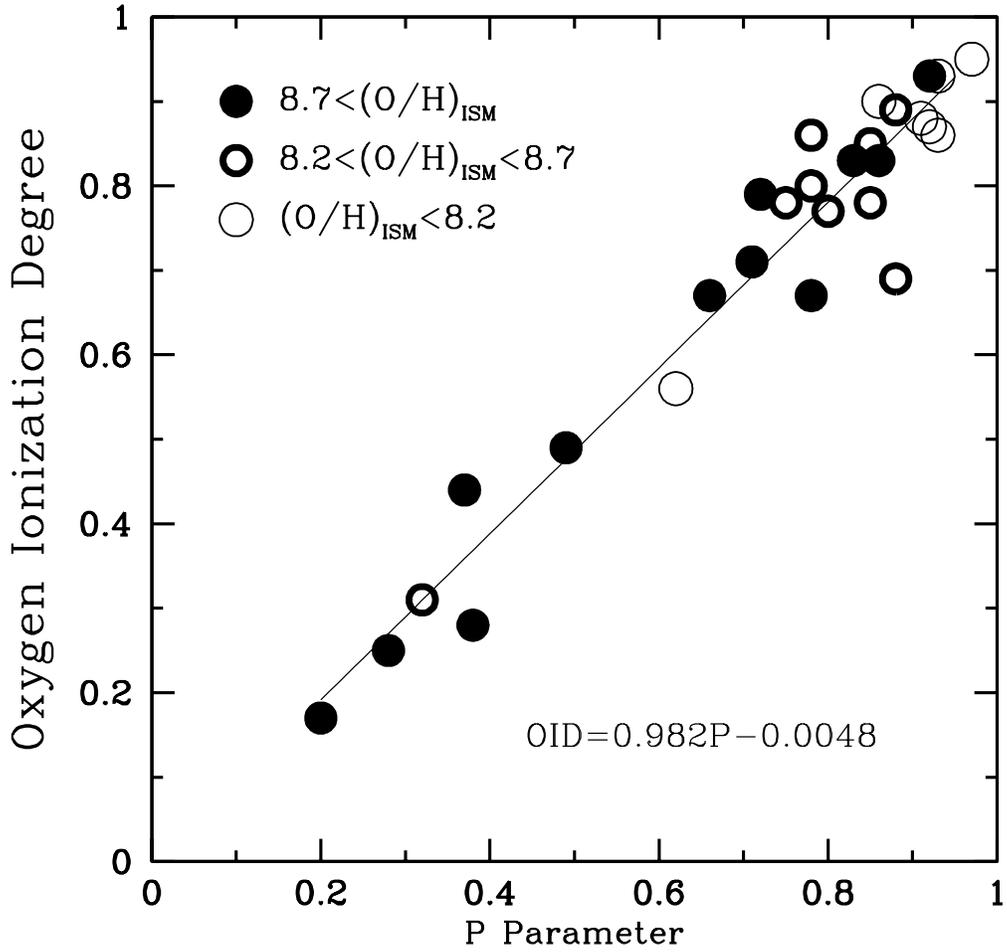}\\
\caption{
Linear relation found between the observational parameter $P$ and the true O$^{++}$ ionization degree. The objects used to obtain this relation are those listed in Table \ref{trec}.
\label{ecoipv3}}
\end{center}
\end{figure}

Since there is a tight relationship between the parameter $P$ and the O$^{++}$ ionization degree (see Figure \ref{ecoipv3}), the fraction of O$^{++}$ in the object can be estimated through the the following equation:
\begin{equation}
\rm OID = 0.9821P-0.0048.
\label{Poid}
\end{equation}
 This equation allows us to statistically estimate the O$^{++}$ ionization degree directly from the observable $P$ without having to make a lengthy analysis of the ionic abundances. It should be noted that while this relation shows no significant bias with the O/H ratio, more points are needed to confirm the lack of bias or to estimate the relevance of the oxygen abundance in this correlation.

\section{Conclusions}\label{conc}

We used a sample of 28 \Hiirs\ obtained from the literature, with \tc\ values ranging from 0.020 to 0.120, to present an important correction to the abundances derived with the \oiiim, as well as a recalibration of Pagel's strong-line method, \odt\, to determine chemical abundances. Both, the correction and the recalibration consider the thermal and ionization structure of the object, as well as the fraction of oxygen depleted into dust grains. We found that the differences between the corrected abundances and those determined with the \oiiim\ lie in the 0.22 to 0.35 dex range.

Almost all previous calibrations of the \odtOH\ in the literature have considered a homogeneous temperature throughout the object. The recalibration presented in equations \ref{recA}\ and \ref{recB}, is the first to simultaneously consider the thermal structure of the photoionized object, the ionization structure, and the correction due to the fraction of oxygen depleted into dust grains. This recalibration is particularly useful for objects with low to intermediate redshifts (up to $z\approx1$).

We find that the O$^{++}$ ionization degree shows a surprisingly almost one-to-one relation with the parameter $P$, thus ensuring that a calibration which considers $P$ is directly considering the oxygen ionization structure of the object.

Given that all strong-line indicators attempt to determine the total oxygen abundance as means of estimating the heavy element content, and that all the observational calibrations have been done using abundances determined with the \oiiim, the oxygen correction presented in this paper should be used to recalibrate other strong-line indicators in the literature.

The up-shifting of the family of curves in the \odtOH\ found in this work, has important consequences in various areas of astrophysics such as the determination of: initial mass functions, star formation rates, luminosity functions, yields, radial metallicity gradients of spiral galaxies, and chemical evolution models of the Universe.

We are grateful to an anonymous referee for a careful reading of the manuscript and several useful suggestions. MAPG and AP received partial support from UNAM (grant PAPIIT 112911) and AP and MP received partial support from CONACyT (grant 129753).

\begin{deluxetable}{lccccccccc} 
\rotate
\tabletypesize{\small}
\tablecaption{\Hiir\ data
\label{trec}}
\tablewidth{0pt}
\tablehead{
\colhead{Object}  & \colhead{Type\tablenotemark{a}}  & \colhead{$T_e$[\ion{O}{3}]} &  \colhead{\tc}  & \colhead{log(\odt)} & \colhead{$P$} &  \colhead{O$^{++}$}& \colhead{O/H\tablenotemark{b}}  & \colhead{O/H\tablenotemark{c}}&\colhead{References}\\
\cline{7-7}
&&&&&&\colhead{O$^+$+O$^{++}$}

}
\startdata
Upper branch\\
NGC 3576	&	GR&	8500$\pm$50	&	0.038$\pm$0.009	& 0.78& 0.78&	0.67	&	8.56	&	8.92	& 1\\
M16			&	GR&	7650$\pm$250&	0.039$\pm$0.006	& 0.59& 0.28&	0.25	&	8.50	&	8.90	& 2\\
M17			&	GR&	8950$\pm$380&	0.033$\pm$0.005	& 0.73& 0.83&	0.83	&	8.52	&	8.88	& 3\\
M8			&	GR&	8090$\pm$140&	0.040$\pm$0.004	& 0.53& 0.38&	0.28	&	8.51	&	8.85	& 3\\
H1013		&	XR&	7370$\pm$630&	0.037			& 0.42& 0.49&	0.49	&	8.45	&	8.84	& 4\\
NGC 595		&	XR&	7450$\pm$330&	0.036			& 0.51& 0.37&	0.44	&	8.45	&	8.80	& 4\\
M20			&	GR&	7800$\pm$300&	0.029$\pm$0.007	& 0.60& 0.20&	0.17	&	8.53	&	8.79	&2\\
Orion		&	GR&	8300$\pm$40	&	0.028$\pm$0.006	& 0.77& 0.86&	0.83	&	8.51	&	8.79	&5, 6\\
NGC 3603	&	GR&	9060$\pm$200&	0.040$\pm$0.008	& 0.89& 0.92&	0.93	&	8.46	&	8.78	& 2\\
K932		&	XR&	8360$\pm$150&	0.033			& 0.72& 0.72&	0.79	&	8.41	&	8.73	&4\\
NGC 2403	&	XR&	8270$\pm$210&	0.039			& 0.59& 0.66&	0.67	&	8.36	&	8.72	&4\\
NGC 604		&	XR&	8150$\pm$160&	0.034$\pm$0.015	& 0.60& 0.71&	0.71	&	8.38	&	8.71	&4\\
S 311		&	GR&	9000$\pm$200&	0.038$\pm$0.007	& 0.72& 0.32&	0.31	&	8.39	&	8.67	& 7\\
NGC 5447	&	XR&	9280$\pm$180&	0.032			& 0.85& 0.78&	0.86	&	8.35	&	8.63	& 4\\
30 Doradus	&	XR&	9950$\pm$60	&	0.033$\pm$0.005	& 0.90& 0.85&	0.85	&	8.33	&	8.61	& 8\\
NGC 5461	&	XR&	8470$\pm$200&	0.027$\pm$0.012	& 0.71& 0.80&	0.77	&	8.41	&	8.60	& 4, 9\\
NGC 5253	&\HG&	11960$\pm$290&	0.072$\pm$0.027	& 0.96& 0.85&	0.78	&	8.18	&	8.56	& 10 \vspace{1.2mm}\\
Transition zone\\
NGC 6822-V	&	XR&13000$\pm$1000&	0.076$\pm$0.018	& 0.91& 0.88&	0.89	&	8.08	&	8.45	&	 11\\
NGC 5471	&	XR&	14100$\pm$300&	0.082$\pm$0.030	& 0.93& 0.75&	0.78 &	8.03 &	8.33 & 9\\
NGC 456  	&	XR&	12165$\pm$200&	0.067$\pm$0.013	& 0.83& 0.78&	0.80	&	7.99	&	8.33	& 12\vspace{1.2mm}\\
Lower branch\\
NGC 346		&	XR&	13070$\pm$50&	0.022$\pm$0.008	& 0.92& 0.88&	0.69&	8.07	&	8.23	& 13, 14\\
NGC 460		&	XR&	12400$\pm$450&	0.032$\pm$0.032	& 0.81& 0.62&	0.56	&	7.96	&	8.19	& 12\\
NGC 2363	&	XR&	16200$\pm$300&	0.120$\pm$0.010	& 0.92& 0.97&	0.97	&	7.76	&	8.14	& 4\\
\told			&\HG&	15800$\pm$170&	0.107$\pm$0.034	& 0.91& 0.92&	0.86	&	7.79	&	8.09	& 15\\
\tolc			&\HG&	14870$\pm$230&	0.029$\pm$0.064	& 0.93& 0.93&	0.87	&	7.90	&	8.12	& 15\\
Haro 29		&\HG&	16050$\pm$100&	0.019$\pm$0.007	& 0.91& 0.91&	0.88&	7.87	&	8.05	& 13, 16\\
SBS 0335$-$052&\HG&	20500$\pm$200&	0.021$\pm$0.007	& 0.67& 0.93&	0.93&	7.35	&	7.60	& 13, 17\\
I Zw 18		&\HG&	19060$\pm$610&	0.024$\pm$0.006	& 0.47& 0.86&	0.90&	7.22	&	7.41& 13, 17\\
\enddata
\tablenotetext{a}{GR=Galactic region, XR=Extragalactic region, \HG= \ion{H}{2} Galaxy.}
\tablenotetext{b}{Gaseous O abundance with homogeneous temperature, \tc=0.000. In units of 12+log(O/H).}
\tablenotetext{c}{Total O abundance with thermal inhomogeneities, \tc$>$0.000, including the correction due to depletion of O into dust grains. In units of 12+log(O/H).}
\\
References.--- (1) \citet{gar04}; (2) \citet{gar06}; (3) \citet{garb07}; (4) \citet{est09}; (5) \citet{est04}; (6) \citet{ode03}; (7) \citet{gar05}; (8) \citet{pea03}; (9) \citet{est02}; (10) \citet{lop07}; (11) \citet{pea05}; (12) \citet{pen12}; (13) \citet{pei07}; (14) \citet{pei00}; (15) \citet{pea12}; (16) \citet{izo97}; (17) \citet{izo99}.
\end{deluxetable} 
\clearpage

\end{document}